\documentclass[aps,prl,twocolumn,amsmath,amssymb]{revtex4} 
\usepackage{bm,graphicx,amsmath}
\usepackage[english]{babel}
\usepackage{latexsym}
\begin{document}  

\title{Lattice solitons in quasicondensates}

\author{V. Ahufinger and A. Sanpera}
\affiliation{Institut f\"ur Theoretische Physik, Universit\"at Hannover, D-30167 Hannover,Germany}

\begin{abstract} 
We analyze finite temperature effects in the generation of 
bright solitons in condensates in optical lattices. 
We show that even in the presence of strong phase fluctuations solitonic structures with well defined phase profile can be created. 
We propose a novel family of variational functions which describe well the properties of these solitons and account for the
nonlinear effects in the band structure. We discuss also the mobility 
and collisions of these localized wave packets.                                       
\end{abstract}  

\maketitle  

Bose-Einstein Condensates (BEC) in optical lattices (OL) are unique candidates 
to explore phenomena that are often extremelly elusive in other areas of physics. 
The dynamics of the system is dominated by the
interplay between the nonlinearity, that can be modified via Feshbach resonances~\cite{Fesh},  
and the periodicity which can be engineered through the intensity, 
geometry, polarization and phase of the two counterpropagating 
laser beams conforming the optical lattice. 
BEC in optical lattices exhibits phenomena 
known from solid state physics as demonstrated for instance
in~\cite{kasevich}. Furthermore, its similarity to other cubic
nonlinear periodic media 
has stimulated a renewed interest in solitonic 
models~\cite{Zobay,nosaltres,malomed1,elenanew,Baizakov}. 

Solitons, which, strictly speaking, are exact solutions 
of integrable models corresponding to wave packets that propagate
without change of their shapes and velocities even in the presence of
collisions, appear in many branches of physics.
In one-dimensional homogeneous condensates with attractive interactions,  
bright solitons exist~\cite{zakarov}, 
and have been recently observed~\cite{Salomon}.   
The presence of a periodic potential, like, e.g., an optical lattice, breaks the
translational invariance, and the system becomes most likely non-integrable, 
having less conserved quantities than degrees of freedom. 
Nonintegrable systems, however, admit also localized solutions which 
are as well commonly termed solitons. These structures differ from proper
solitons either in their motion and/or their collisions. 
A well known example are optical solitons in periodic media 
(cf. \cite{reviewopt}) whose interactions 
have been extensively studied  
using either 
the discrete nonlinear Schr\"{o}dinger (DNLS) equation \cite{Aceves} 
or weakly perturbed integrable models~\cite{Semagin}.

So far, to our knowledge, the generation of lattice solitons in
condensates has been only discussed at zero temperature 
~\cite{Zobay,nosaltres,malomed1,elenanew,Baizakov}, where there is 
an analogy between an array of optical waveguides in a Kerr medium 
and a BEC in a periodic potential. 
In this limit, one can generate \cite{Oberthaler2} bright lattice solitons 
in repulsive condensates when the tunneling rate balances 
the nonlinear energy of the system. This compensation occurs if the soliton is placed at the edge of 
the first Brillouin zone where the effective mass 
becomes negative reaching the staggered 
configuration~\cite{staggered}.

Here we discuss the generation of solitons in repulsive condensates
at finite temperature. We show that 
even  when the condensate is not phase coherent,
solitonic structures with well defined phase profile can be created. 
A new insight into the nature of these lattice solitons is obtained 
by means of a novel variational ansatz that accounts 
for the effects of the nonlinearity. 
Finally, we address the issue of mobility and collisions.

For BEC in 3D trapping geometries, fluctuations 
of density and phase are only important in a 
narrow temperature range near the BEC transition 
temperature $T_c$~\cite{Petrov2}. 
For pure 1D systems, however, phase fluctuations are present at 
temperatures far below the degeneracy temperature  
$T_d=N\hbar \omega_x/k_B$, 
while density fluctuations are still 
suppressed~\cite{Petrov1} 
($k_B$ denotes the Boltzmann constant, $N$ the number
of atoms, $\omega_x$ the axial trapping frequency).
Phase fluctuations can be studied 
by solving the Bogoliubov-de
Gennes (BdG) equations describing elementary excitations. 
Writing the quantum field operator as  
$\hat{\psi}(x)=\sqrt{n_0(x)}\exp(i\hat{\phi}(x))$ where  $n_0(x)$
denotes the density, the phase operator takes the form \cite{Shevchenko}:
\begin{equation}
\hat{\phi}(x)=\frac{1}{\sqrt{4 n_0(x)}}
\sum_{j=1}^{\infty}f_{j}^{+}(x)\hat{a}_j+H.c.
\end{equation}
where $\hat{a}_j$ is the annihilation operator of the excitation with
quantum number $j$ and energy 
$\epsilon_j=\hbar \omega_x \sqrt{j(j+1)/2}$, and  $f_{j}^{+}=u_j+v_j$,
where $u_j$ and $v_j$ denote the excitation functions determined by
the BdG equations. In 1D and in the Thomas-Fermi (TF) regime, 
the functions $f_{j}^{+}$ have the form:
\begin{equation}
f_{j}^{+}(x)=\sqrt{\frac{(j+1/2)2\mu}{R_{TF}\epsilon_j}
\left(1-\left(\frac{x}{R_{TF}}\right)^2\right)} P_j\left(\frac{x}{R_{TF}}\right)
\end{equation}
where $ P_j(x/R_{TF})$ are Legendre polynomials,
$R_{TF}=(2\mu/m\omega_x{}^2)^{1/2}$ is the TF radius and $\mu$ the chemical potential.
The phase coherence length, $L_{\phi}=R_{TF}T_d\hbar \omega_x/\mu T$, characterizes the maximal distance
between two phase correlated points in the condensate. Phase
fluctuations increase for large trap 
aspect ratios, $\omega_t/\omega_x$, and small $N$ \cite{Dettmer}. 

To study temperature effects in the generation of lattice solitons, 
we consider a $^{87}$Rb condensate with $N=500$ atoms in a magnetic trap 
with frequencies $\omega_t=715\times2\pi$ Hz and
$\omega_x=14\times2\pi$ Hz. 
For such parameters, the system is effectively in a 1D regime 
since transverse excitations are supressed ($\mu<<\hbar \omega_t$).
Moreover, along the axial direction, the condensate is well in the TF
regime ($\mu>>\hbar \omega_x$).
To create lattice solitons in repulsive condensates,
the optical lattice depth  belongs to the so-called weak
potential regime. Thus, one cannot use the tight binding approximation~\cite{mermin}
to rewrite the condensate order parameter as a sum of wavefunctions
localized in each well of the lattice.
To study the dynamics of the system we use
the 1D Gross Pitaevskii equation (GPE):
\begin{equation}
i\hbar \dot{\psi}(x,t) = \left(-\frac{\hbar^2}{2m}\nabla+
V \left( x ,t\right)+g\vert \psi\left( x,t \right)\vert^2\right)\psi
\left( x,t \right),
\label{GPE}
\end{equation}
with coupling constant $g=2 N \hbar a_s \omega_{t}$ and 
scattering length  $a_s=$5.8 nm.
The confining potential 
$V (x,t)= m\omega_x(t)^2 x^2/2+ V_0(t)\sin^2(\pi x/d)$
describes both the axial magnetic trap 
and the optical lattice. The later is characterized by its maximal  
depth $V_0=1 E_r$ and by its spatial period $d=\lambda/2$
($\lambda=795$nm). Energy is expressed in units of the recoil energy 
$E_r=\hbar^2 k^2/2m$, where $k=\pi/d$.
Temperature is included at the level of the GPE
by calculating first the density at $T=0$ in the presence of the
magnetic trap only, i.e., $V(x,t=0)=m\omega_x^2 x^2/2$   
and then the phase operator (1). To this aim,
we calculate the Bose ocuppation $N_j=[e^{\epsilon_j/k_B}-1]^{-1}$ 
of the low-energy modes in the Bogoliubov approximation for a fixed T
replacing the operators $\hat{a}_j$ and $\hat{a}_j^{+}$ 
by random complex variables $\alpha_j$ and
$\alpha_j^{*}$ such that $\langle|\alpha_j|^2\rangle=N_j$
\cite{Dettmer}. One can also study the quasi 1D case using a 3D GPE and 
replacing in (1) the energy of low excitations by $\epsilon_j=\hbar
\omega_x \sqrt{j(j+3)/4}$ \cite{Stringari} 
and the Legendre by Jacobi polynomials~\cite{Dettmer}.

We summarize the procedure to generate lattice solitons at a given $T$.
Once the phase fluctuations have been included, 
periodicity is introduced by growing adiabatically 
the optical lattice (t$\le \tau$). 
During this process, density fluctuations arise. We simultaneously turn off the magnetic trap 
and imprint a phase difference of $\pi$ between
consecutive wells to place the condensate at the edge of the first
Brillouin zone. This is achieved by using an 
auxiliary optical lattice with double spatial period than the first
one, acting for a time much shorter than the tunneling and the
correlation time  of the system ~\cite{nosaltres}. Hence, the condensate acquires the desired phase without modifying its
density profile.
\begin{figure}
\includegraphics[width=1.05\linewidth]{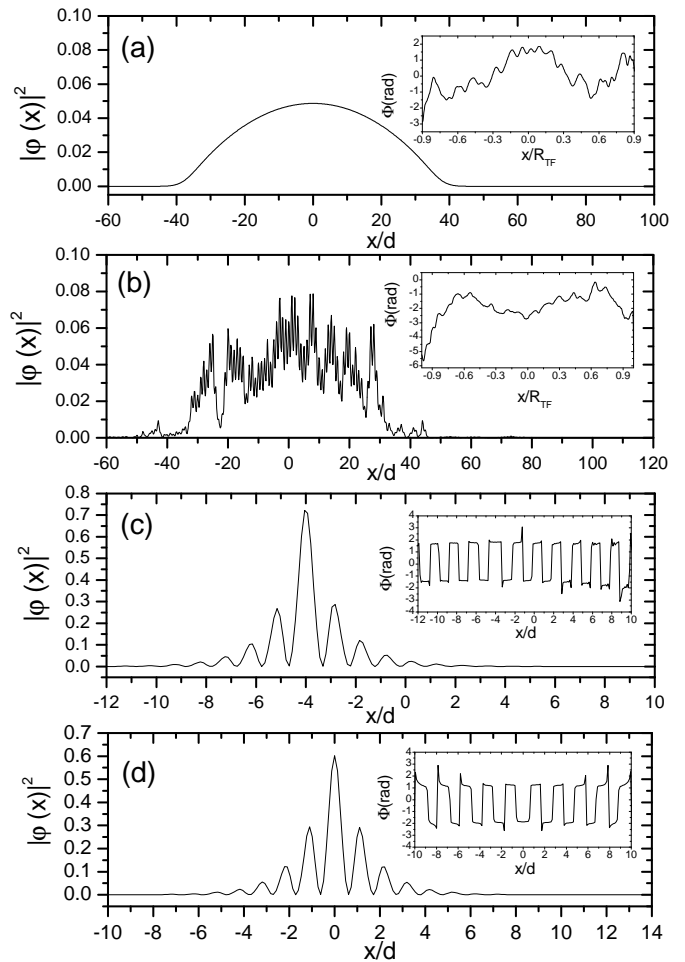}
\caption{Density and phase (inset) profile of (a) magnetically trapped ($\omega_x=14\times
  2\pi$ Hz) ${^{87}}$Rb ground state condensate with $N=500$ at $T=0.8T_c$, 
(b) after the adiabatic growing of the optical lattice ($V_0=1E_r$ and $\lambda=795$nm), (c) of lattice soliton 100ms after the imprinting of a phase difference 
of $\pi$ between  consecutive wells and after the magnetic trap is switched off
  (d)of lattice soliton generated under the same conditions of (c) but at $T=0$.}\label{fluct}
\end{figure}

Figure \ref{fluct} shows density and phase of the condensate at $T=0.8T_c$ ($T_c=N\hbar\omega_x/k_B ln{2N}$ \cite{Ketterle}) (a) 
in the presence of the magnetic trap only ($t=0$),
(b) after growing adiabatically the optical lattice ($t=\tau$), 
and (c) 100ms after the phase imprint has been performed and the magnetic
trap has been switched off ($t=\tau+100$). Figure 1(d) shows a lattice soliton generated at $T=0$ 
with otherwise identical parameters.
Despite the strong fluctuations of phase and
density induced by finite temperature, the system (after phase imprint) evolves towards a 
staggered soliton configuration. It contains
approximately 35\% of the initial atoms 
and remains so for times much larger than the tunneling time.
This robustness can be understood by realizing that the size of the lattice soliton 
is smaller than the phase coherence length for all $T<T_c$. 
The soliton size (which depends on N, being smaller
as N increases ) is independent of the temperature 
but due to the random character of the fluctuations 
the position of the generated soliton is different for each realization. 

An analytical description of lattice solitons in repulsive condensates 
is often performed through an effective theory for the soliton's 
envelope in the effective mass approximation~\cite{Steel}. 
The influence of the periodic potential is 
included there via an effective mass and coupling constant. Instead let us consider a very general ansatz 
for the soliton wavefunction: 
\begin{equation}
\psi(x,t)=e^{-i\mu t/\hbar}G(A,\sigma,x,x_0)F(x)
\label{Ansatz}
\end{equation} 
where $G(A,\sigma,x,x_0)$ describes the envelope of a soliton 
of amplitude A, width $\sigma$, and centered at $x_0$. 
The effect of the periodic potential is included, in the weak approximation limit, 
as a combination of linear Bloch functions $F(x)=\sum_{k}e^{ikx}f_k(x)$, where the $f_k(x)$ have the periodicity of the lattice.
At the edge of the first Brillouin zone, this combination can be
approximated by only two harmonics, i.e., $F(x)= \cos(2\pi x/\lambda)$
~\cite{malomed1}. Notice that such an ansatz does not take into
account nonlinear effects. Moreover, it does not present
a minimum in the energy functional of the GPE for any atom number.
Inspection of
Fig.\ref{fluct}(c) and (d) shows that the density profile
inside each well is shifted with respect to the minimum
of the optical potential. A relative simple function that reproduces
such site dependent shift is $F(x)=\cos(2 \pi x/ \lambda'(x))$ with an
effective wavelength
$\lambda'(x)=\lambda(1+\alpha(N)|G(A,\sigma,x,x_0)|^2)$ where
$\alpha(N)<<1$ is a variational parameter
which depends only on $N$. In other words, the modification of
the band structure due to the
nonlinearity results in an effective change of the periodicity of the
system. The ansatz (\ref{Ansatz}) with the effective wavelength $\lambda'(x)$ does not allow a fully analytical treatment for
the energy functional. As a first approximation we assume thus that the
effect of the nonlinearity is to shift $\lambda$ simply by a constant
$\lambda'=\lambda+\delta$. With this new ansatz, 
imposing the normalization of the wavefunction
(i.e., conservation of $N$) and assuming a gaussian
envelope $G(A,\sigma,x,x_0)=A \exp((x-x_0)^2/2 \sigma^2)$ the energy
functional $E[A,\sigma, x_0,\delta]$ reads:
\begin{eqnarray}
&&E=\int\left[\frac{\hbar^2}{2m}|\nabla \psi(x)|^2 +\frac{g}{2}|\psi(x)|^4 +V(x)|\psi(\
x)|^2\right]dx=\nonumber \\
&&B \Bigg( \frac{\hbar^2}{m}\left[\frac{ 1+e^{-{k'}^2 \sigma^2}\cos(2{k'}x_0)}{2\sigma\
^2}+{k'}^2 \right] \nonumber \\
&&+\frac{g|A|^2}{4\sqrt{2}}\left[3+e^{-2 k'^2 \sigma^2}\cos(4k'x_0)+4e^{-\frac{k'^2 \sigma^2}{2}}\cos(2k'x_0)\right]\nonumber \\
&&+V_0\Big[1+e^{-k'^2 \sigma^2}\cos(2k'x_0)-e^{-k^2 \sigma^2}\cos(2k x_0) \nonumber \\
&&  -\frac{1}{2}e^{-k_-^2 \sigma^2}\cos(2k_- x_0) -\frac{1}{2}e^{-k_+^2 \sigma^2}\cos(2k_+ x_0)\Big]
\Bigg).
\label{energies}
\end{eqnarray}
where $B=|A|^2 \sqrt{\pi}\sigma/4$, $k'=2\pi/\lambda'$ and $k_{\pm}=k' \pm k$.
We minimize (\ref{energies}) for a soliton centered in one lattice
site i.e., $x_0=0$, as a function of the shift $\delta$ and the width
$\sigma$. The other free parameter $A$, the
amplitude, is fixed by normalization, i.e., $1\equiv 2B(1+e^{\
-k'^2 \sigma^2}\cos(2k'x_0))$.

For each $N$ and $x_0$, minimization of $E[A,\sigma,x_0,\delta]$ with
respect to  $\sigma$ and $\delta$ presents now a clear minimum. The
values of $\sigma$ and $\delta$
obtained through minimization coincide with the values of the width and
on-site shift obtained by solving numerically the
GPE (\ref{GPE}). Also, minimization of (\ref{energies})
correctly indicates that solitons containing few atoms ($N<100$) are
very extended (large $\sigma$) and present practically no shift
($\delta \leq 0.001$), as one should expect in the continuum limit.
On the contrary for larger $N$, $\sigma$ decreases and $\delta$ grows.

The Peierls-Nabarro (PN) barrier~\cite{PN,nosaltres} 
i.e., the energy difference
between a solitonic configuration centered in one minimum of the
lattice, $E[A,\sigma,x_0=nd,\delta]$, and the one centered in 
one maximum, $E[A',\sigma',x_0=n d/2,\delta']$,
has been also calculated using (\ref{energies}). 
For any value of $N$, $E[A,\sigma,x_0=nd,\delta]< E[A',\sigma',x_0=n d/2,\delta']$. Thus, even the constant shift approximation reproduces the main features of lattice solitons.  
\begin{figure}
\includegraphics[width=1.0\linewidth]{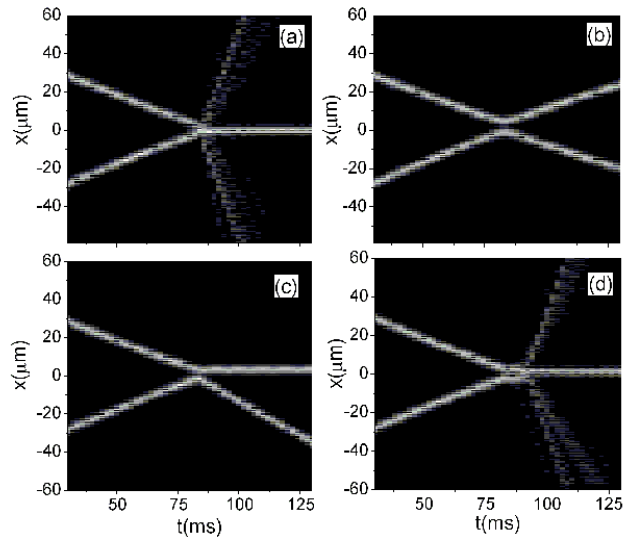}
\caption{Collisions between two identical lattice solitons ($N=200$)
 at an initial distance of $2x_0=246d$ and with an initial phase
difference of (a) $\Delta \phi=0$rad, (b) $\Delta \phi=\pi/2$rad,
(c) $\Delta \phi=0.3$rad and (d) $\Delta \phi=0.9$rad.
The initial transfer of momentum is $0.1k\hbar$ in all the cases.}\label{colis}
\end{figure}

On-site lattice solitons are created initially at rest.
Nevertheless, if they are well localized in momentum space and
an instantaneous transfer of momentum  (large enough to overcome the
PN barrier) is given to the system, the solitons move in opposite
direction to the given momentum due to their negative effective mass.
Notice that the variational ansatz $F(x)=cos(2\pi x/\lambda')$ with
$\lambda'=\lambda+\delta$ is meaningful only in the static case. Such a
simplified solution cannot be used to study soliton dynamics,
since the soliton is always chirped with respect its center.
The Euler-Lagrange equations for an ansatz whose
periodicity depends also on the density are quite complex.
Therefore, to study dynamical behaviour one has to
rely on numerical simulations.

Giving momentum to the system is generally accompanied by radiation of
atoms. The larger the given momentum is the larger the losses.
For small initial transfer of momentum the system exhibits a nonlinear
response, its velocity slows down, and eventually a complete halt
of the soliton occurs.
The given momentum $p$, for which the soliton experiences
a nonlinear response lies on the range $0\leq  p\leq \hbar(k-k')$ where $k'$ corresponds to
the inverse of the effective wavelength.
Thus the linear response is recovered for broad solitons
since $k\simeq k'$ and for well localized solitons if the initial
given momentum is large enough, i.e. $p>\hbar(k-k')$.
Recently nonlinear movement of discrete solitons has
also been reported \cite{malomed1}.
The interaction between lattice solitons depends also strongly on the
initial given momentum. We study numerically collisions between
two identical solitons with $N=200$ each, created initially at rest
and separated by a distance $2x_0=246d$.
For small values of the initial momentum $p\simeq 0.1k \hbar$,
the two initial solitons merge at the interaction point resulting
in a single soliton (with $N=200$) at rest.
The excess of atoms is lost by radiation in a
symmetric way (Fig. 2(a)). For intermediate values of the initial
kick, $p\simeq 0.2k \hbar$,
the two solitons merge in a moving soliton with smaller $N$
than the initial ones.
For larger kicks, $p\simeq 0.3k \hbar$, the solitons exhibit a
quasi elastic collision where the two solitons pass each other.
The average phase difference between the two solitons also affects
the nature of their interactions.
For $p=0.1k \hbar$ and identical initial phase distribution
one obtains the fusion of
the initial two solitons into a single one at rest (Fig. 2(a)). This
merging behaviour is always present
for average phase difference between the solitons below $0.19$rad,
although the final structure can either be a soliton or a breather and
its position can also change.
For phase differences of the order of $\pi/2$ or larger, the two
solitons always repel each other (Fig. 2(b)). Between these two
extreme cases, the dynamics of the system is unpredictable. Two
solitons either with approximately the same $N$ and same
velocity or with different $N$ (being the one with more
atoms the slower) (Fig. 2(c)) are examples of possible situations
after the collision. In other cases, the two initial solitons
form a bound state that ends up into a single soliton
(Fig. 2(d)). These results evidence the strongly dependence on the
initial conditions in the dynamics of the system.

Summarizing, finite temperature effects in the generation of lattice 
solitons in quasicondensates have been addressed. 
New insight into the nature of these structures is
provided through a new family of variational functions in which the
effect of the nonlinearity is shown as an effective change in the
lattice periodicity. We hope that such novel ansatz opens new
possibilities in the study of lattice solitons.

We thank M. K. Oberthaler, E. Ostrovskaya and M. Lewenstein for stimulating discussion\
s. We acknowledge support from Deutsche Forschungsgemeinschaft (SFB 407) and from the \
ESF Programme QUDEDIS. V.A. acknowledges support from the European Community (MEIF-CT-2003\
-501075).


\begin{references}
\bibitem{Fesh}S. Inouye {\it et al.}, Nature (London) {\bf 392},
151 (1998); S. L. Cornish {\it et al.}, Phys. Rev. Lett. {\bf 85}, 1795 (2000).

\bibitem{kasevich}B.~P.~Anderson and M.~A.~Kasevich, Science {\bf 282}, 1686 (1998); 
O.~Morsch {\it et al.}, Phys. Rev. Lett. {\bf 87}, 140402 (2001); 
F.~S.~Cataliotti {\it et al.}, Science {\bf 293}, 843 (2001).

\bibitem{Zobay}O. Zobay {\it et al.}, Phys. Rev. A {\bf 59}, 643 (1999); 
A.~Trombettoni and A.~Smerzi, Phys. Rev. Lett. {\bf 86}, 2353 (2001); 
F.~Kh.~Abdullaev {\it et al.}, Phys. Rev. A {\bf 64}, 043606 (2001); 
E.~A.~Ostrovskaya and Y.~S.~Kivshar, Phys. Rev. Lett. {\bf 90}, 160407 (2003).

\bibitem{nosaltres}V.~Ahufinger {\it et al.}, Phys. Rev. A {\bf 69}, 053604 (2004).

\bibitem{malomed1}M. J. Ablowitz, Z. H. Musslimani and G. Biondini, Phys. Rev. E {\bf 65}, 026602 (2002). 
H.~Sakaguchi and B.~A.~Malomed, J. Phys. B: Mol. Opt. Phys.{\bf 37}, 1443 (2004).

\bibitem{elenanew} B.~J.~ Dabrowska, E.~ A.~Ostrovskaya and Y.~S.~Kivshar, cond-mat/0408234 (2004).

\bibitem{Baizakov}B. B. Baizakov, V. V. Konotop and M. Salerno, J. Phys. B: At. Mol. Opt. Phys. {\bf 35}, 5105 (2002); 
B. B. Baizakov, B. A. Malomed and M. Salerno, Europhys. Lett. {\bf 63}, 642 (2003); 
H.~Sakaguchi and B.~A.~Malomed, J. Phys. B: Mol. Opt. Phys. {\bf 37}, 2225 (2004); 
A. M. Dudarev, R. B. Diener and Q. Niu, J. Opt. B: Quantum Semiclass. Opt. {\bf 6}, S231 (2004).

\bibitem{zakarov}V. E. Zakharov and A. B. Shabat, Sov. Phys. JETP {\bf 37}, 823(1973).

\bibitem{Salomon}L.~Khaykovich {\it et al.}, Science {\bf 296}, 1290 (2002); 
K.~E.~Strecker {\it et al.}, Nature {\bf 417}, 150 (2002).

\bibitem{reviewopt} Y. S. Kivshar and G. Agrawal, {\it Optical Solitons}, Academic Press 2003.

\bibitem{Aceves}A.~B.~Aceves {\it et al.}, Phys. Rev. E {\bf 53}, 1172 (1996); 
D.~Cai, A.~R.~Bishop and N.~Gr$\o$nbech-Jensen, Phys. Rev. E {\bf 56}, 7246 (1997); 
M.~J.~Ablowitz and Z.~H.~Musslimani, Phys. Rev. E {\bf 65},056618 (2002); 
I.~E.~Papacharalampous  {\it et al.}, Phys. Rev. E {\bf 68}, 046604 (2003).

\bibitem{Semagin}D.~A.~Segamin {\it et al.}, Physica B {\bf 316}, 136 (2002); 
S.~V.~Dmitriev {\it et al.}, Phys. Rev. E {\bf 66}, 046609 (2002); 
S.~V.~Dmitriev and T.~Shigenari, Chaos {\bf 12}, 324 (2002); 
S.~V.~Dmitriev {\it et al.}, nlin.PS/0309054 (2003).

\bibitem{Oberthaler2} B.~Eiermann {\it et al.}, Phys. Rev. Lett. {\bf 92}, 230401 (2004).

\bibitem{staggered} Y.~S.~Kivshar, Opt. Lett. {\bf 18}, 1147 (1993); 
J. W. Fleischer  {\it et al.}, Phys. Rev. Lett. {\bf 90}, 023902 (2003).

\bibitem{Petrov2}D.~S.~Petrov, G.~V.~Shlyapnikov, and J.~T.~M.~Walraven, Phys. Rev. Lett. {\bf 87}, 050404 (2001).

\bibitem{Petrov1}D.~S.~Petrov, G.~V.~Shlyapnikov, and J.~T.~M.~Walraven, Phys. Rev. Lett. {\bf 85}, 3745 (2000).

\bibitem{Shevchenko}S.~I.~Shevchenko, Sov. Low. Temp. Phys. {\bf 18}, 223 (1992); 
H.~Kreutzmann {\it et al.}, Appl. Phys. B {\bf 76}, 165 (2003).

\bibitem{Dettmer}S.~Dettmer {\it et al.}, Phys. Rev. Lett. {\bf 87}, 160406 (2001).

\bibitem{mermin} N. W. Ashcroft and N. D. Mermin {\it {Solid State Physics}}, Saunders College (1976).

\bibitem{Stringari}S. Stringari, Phys. Rev. A {\bf 58}, 2385 (1998).
\bibitem{Ketterle}W.~Ketterle and N.~J.~ van Druten, Phys. Rev. A {\bf 54}, 656 (1996).
\bibitem{Steel}H. Pu {\it et al.}, Phys. Rev. A {\bf 67}, 043605 (2003).
\bibitem{PN}Y.~S.~Kivshar and D.~K.~Campbell, Phys. Rev. E {\bf 48}, 3077 (1993) and references therein.
\end{references}
\end{document}